\documentclass[conference]{IEEEtran}
\IEEEoverridecommandlockouts
\usepackage{cite}
\usepackage{amsmath,amssymb,amsfonts}
\usepackage{algorithmic}
\usepackage{graphicx}
\usepackage{textcomp}
\usepackage{xcolor}
\usepackage{multirow}
\usepackage{hyperref}

\hypersetup{
    colorlinks=true,
    citecolor = black,
    linkcolor=black,
    filecolor=black,      
    urlcolor=black,
}

\def\BibTeX{{\rm B\kern-.05em{\sc i\kern-.025em b}\kern-.08em
    T\kern-.1667em\lower.7ex\hbox{E}\kern-.125emX}}

\begin{document}

\title{\huge \textbf{Densely Connected Residual Network for Attack Recognition\\ \thanks{Peilun Wu\IEEEauthorrefmark{1} and Nour Moustafa\IEEEauthorrefmark{2} are equally contributed to this work. Free use of the Ton\_IoT dataset for academic purposes is hereby granted in perpetuity. Use for commercial purposes is allowable after asking the author, Dr Nour Moustafa, who has asserted his right under the copyright.} }}

\author{\IEEEauthorblockN{Peilun Wu\IEEEauthorrefmark{1}, Nour Moustafa\IEEEauthorrefmark{2}, Shiyi Yang\IEEEauthorrefmark{3} and Hui Guo\IEEEauthorrefmark{4}}
\IEEEauthorblockA{School of Computer Science and Engineering, University of New South Wales, Sydney\IEEEauthorrefmark{1}\IEEEauthorrefmark{3}\IEEEauthorrefmark{4}\\
Australian Center for Cyber, University of New South Wales, Canberra\IEEEauthorrefmark{2}\\
Innovation Institute, Sangfor Technologies Inc.\IEEEauthorrefmark{1}}
Email: \IEEEauthorrefmark{1}z5100023@zmail.unsw.edu.au,
\IEEEauthorrefmark{2}nour.moustafa@unsw.edu.au, \\
\IEEEauthorrefmark{3}z5223292@cse.unsw.edu.au,
\IEEEauthorrefmark{4}h.guo@unsw.edu.au}

\maketitle

\begin{abstract}
High false alarm rate and low detection rate are the major sticking points for unknown threat perception.
To address the problems, in the paper, we present a densely connected residual network (Densely-ResNet) for attack recognition.
Densely-ResNet is built with several basic residual units, where each of them consists of a series of Conv-GRU subnets by wide connections. 
Our evaluation shows that Densely-ResNet can accurately discover various unknown threats that appear in edge, fog and cloud layers and simultaneously maintain a much lower false alarm rate than existing algorithms.

\end{abstract}


\section{Introduction}\label{introduction}
It is well-known that machine learning (ML) has prompted the development of information technology in many aspects, such as face recognition and machine translation.
It is no exception that ML can also be used for attack recognition to improve the trustworthiness and dependability of computer systems.
However, attack recognition is known as a task of \textbf{\textit{``finding very damaging needles in very large haystacks"}}\cite{sommer2010outside}, which exhibits disparate difficulties and presents great challenges in contrast to other classification tasks when using machine learning.  

\begin{figure}[t]
    \centering
    \includegraphics[width=.85\linewidth]{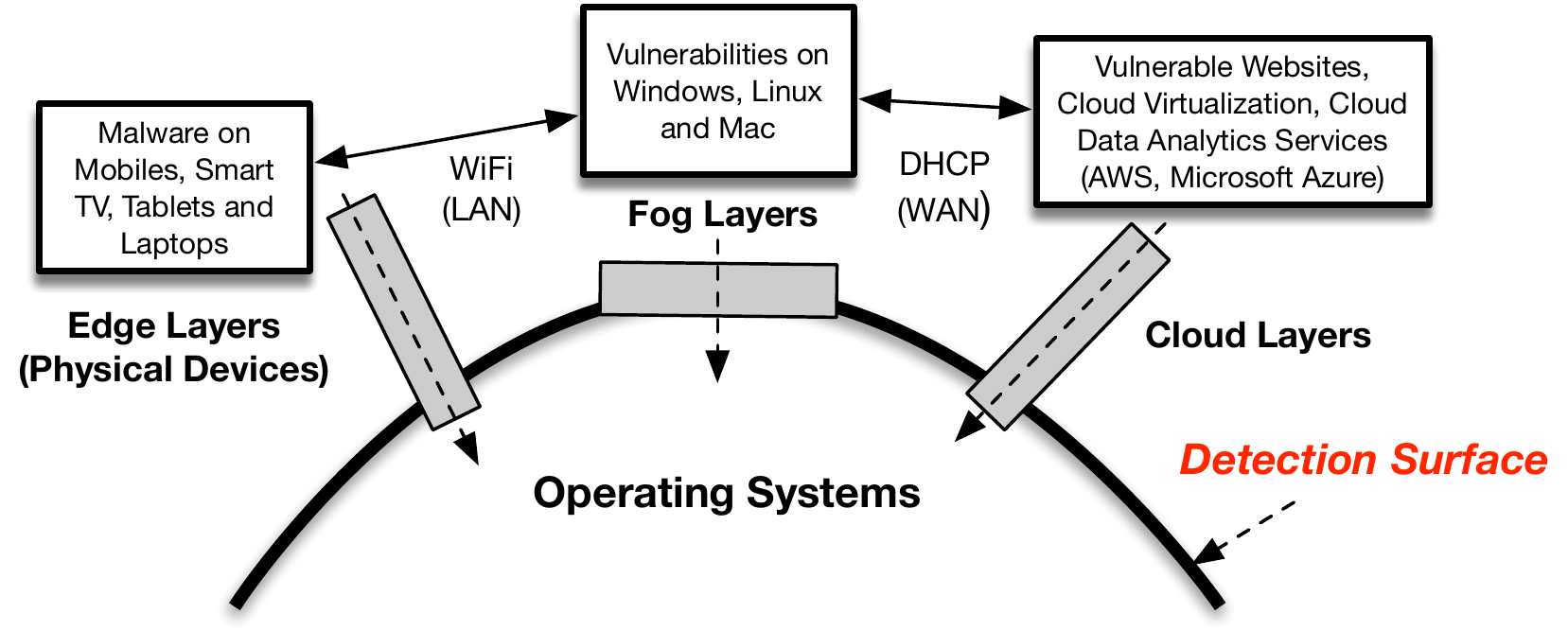}
    \caption{Detection Surface}
    \label{fig:detection_surface}
\end{figure}

We claim that using ML for attack recognition should be careful, which must be accompanied with an elaborately defined threat model for the sake of minimizing the scale of detection surface. 
Nevertheless, due to the monotonous attack spread-ability, traditional detection surface\cite{denning1987intrusion} that is defined based on abnormal patterns of system usage cannot be fully adapted and representative for the status in quo of security requirement.
Based on the principle of \textbf{\textit{``defense in depth"}}\cite{lippmann2006validating}, we propose an alternative detection surface, which is dependent on data fusion from edge layers to cloud layers as shown in Fig~\ref{fig:detection_surface}. 
The new detection surface connects three virtual layers (edge, fog and cloud layers) as a whole in order that a defense system can be more comprehensively to discover external attacks that attempt to invade an operating system but from different dimensions.

In this paper, a densely connected residual network (Densely-ResNet) is deployed on the detection surface, which is constructed with several basic residual units, where each of them contains two Conv-GRU subnets by wide connections.
Densely-ResNet has several compelling advantages: it alleviates the vanishing-gradient problem, strengthens the learning stability, reduces the feature loss, encourages the feature reuse and extends the depth and width of neural network in a feed-forward fashion.
The evaluation result shows that Densely-ResNet tends to increase the complexity of computation but demonstrates a much better accuracy and lower false-positive rate than existing algorithms for attack recognition and significantly outperforms the current state-of-the-art results on the benchmark attack dataset.
Our contributions are summarized as follow:
\begin{itemize}
    \item We construct a new cybers threat assessment testbed called Ton\_IoT that involves Edge (IoT devices), Fog (endpoints) and Cloud (network services) layers as a syncretic detection surface, which can be used to monitor and collect multi-sourced system records from various attack scenarios.
    \item We propose a novel densely connected residual network (Densely-ResNet) for attack recognition and unknown threat perception. Densely-ResNet can achieve state-of-the-art detection accuracy and lowest false alarm rate on UNSW-NB15 benchmark, which significantly outperforms existing algorithms. 
    \item We develop a correlation analysis module to post-analyze the prediction result of Densely-ResNet, which can remove the redundant alarms that are not false positives but unsuccessful attack attempts to mitigate the problem of alarm fatigue.
\end{itemize}

\section{Background and Related Work}
\subsection{Attack Classification}
\subsubsection{Reconnaissance}
Before beginning to breach a victim machine, hackers commonly utilize tools such as Nessus \cite{Nessus} and Nmap \cite{Nmap} to scan and sniffer the local network configurations of the targeted system, which aims to discover some available ports that are opened and can be potentially exploited.
Besides, after the stage of privilege escalation, hackers may sign into the victim machine and then execute a series of commands to run an internal reconnaissance, such as DNS Reconnaissance ($cmd \rightarrow nslookup$), Directory Service Reconnaissance ($cmd \rightarrow net$  $user/domain$) and User and IP Address Reconnaissance ($cmd \rightarrow NetSess.exe$ $ContosoDC$), which purposes to discover computers of interest, enumerate users and groups, collect important IP addresses, and evaluate the organization's assets and weaknesses.
\subsubsection{SQL Injection (SQLI) Attack}
SQLI allows malicious users to execute illegal SQL quires that are embedded with URL parameters to modify the database for extracting valuable and sensitive information.
Complex variations of SQL injection attack, such as `Out-of-Band SQLI', can also be launched by phishing emails that consist of the query results with injected SQL codes, which can result in data theft, data loss and even full system compromise. 

\subsubsection{Cross-Site Scripting (XSS) Attack}
XSS can inject arbitrary JavaScript into a legitimate website or web application which is then executed inside a victim’s browser.
The JavaScript contains malicious codes, such as webshell or `one word Trojan', once being executed, it will build a network socket to communicate with the offensive server.
White/Black list strategy is a common defense skill to prevent the JavaScript injection; however, because of the misused and insecure JavaScript programming, XSS attacks are still active in common attack categories and existing many skills to bypass the white/black list detection mechanism. 
\subsubsection{Distributed Denial of Service (DDoS)}

DDoS attack requires to remotely control a number of infected computers (Botnets) then command them to send a massive volume of legitimated requests through a variety of protocols, such as HTTP flood, SYN flood and DNS amplification, to overflow the capacity of a targeted system. 
The goal of DDoS attack is attempting to congest the traffic result in service disable.

\subsubsection{Ransomware and Worm}
Ransomware attack is a form of malware that can encrypt victim's files and then demand a ransom from the victim to restore access to the data upon payment.
Ransomware are typically carried out using a Trojan that is disguised as a legitimate file that the user is tricked into downloading or opening when it arrives as an email attachment.

\subsubsection{Advanced Persistent Threat (APT)}

APT is a combination of several elaborately designed cyber attacks, which use different tactic techniques launched by well-trained hackers who aim to steal sensitive and valuable information from enterprises and governments out of different profitable purposes.
APTs can utilize a combination of vulnerabilities to exploit the target systems for building the Command \& Control (C\&C) communications between several victim machines and offensive systems \cite{milajerdi2019holmes}. 

Due to the high degree of their sophistication and concealing, APTs are hard to be discovered by traditional intrusion detection systems, of which detection engines are based on association rules and abnormal patterns.
Besides, a lack of relatively real and sophisticated threat assessment environment is also one of the serious problems for the experimental evaluation of APT perception. 

Furthermore, an APT may also take advantages of a variety of break-in methods, such as driven-by-download or a spear-phishing attack, which will invoke a series of processes and I/O operations each time that demonstrates a significant diversification against the assumption that we naively proposed for the trace detection - \textbf{\textit{``an APT may follow an unique file path differed from normal behaviour in a provenance graph"}}\cite{ma2016protracer}.
It can be argued that a hacker may have various approaches to break in a system, but whom may only have a few ways to clean up the attack remnants and the consequences of intrusion behaviour are irreversible. 

Thus, we suggest that, except the analysis of file paths based on their corresponding provenance graphs, the result-oriented detection is also essential for APT detection, which should be included within the kernel-level audit data, such as the occurrences of I/O operation (read/write file), the changes of virtual memory allocation in bytes (execute malicious bash script), the incremental sizes of outbound network packet (build C\&C communication) and the performance consumption of CPU processor (trace elimination / side channel attack).
It can be assumed that, when an APT is active in a victim system, it may present kernel-level characteristic differences compared to the normal usage of system activities.

\subsection{Defense Mechanism}
\subsubsection{Signature-Based Pattern Matching Defense System}

The most common-used intrusion detection system (IDS) in commercial environments is signature-based or rule-based.
Blue team can simulate attacks in a controlled test environment and then extract the most representative features as attack signatures through an in-depth runtime analysis. 
IDS can make relevant association rules in line with well-defined attack signatures to construct a black list to prevent both internal and external intrusions.
It has been proved that the signature-based or rule-based IDS can achieve a low false alarm rate (FAR) in practice; however, the weakness is that the hand-designed attack signatures contains excessively detailed descriptions of some specific attacks, which make IDS incapacitate to discover novel attacks \cite{hubballi2014false}.

\begin{figure*}[t]
    \centering
    \includegraphics[width=.88\linewidth]{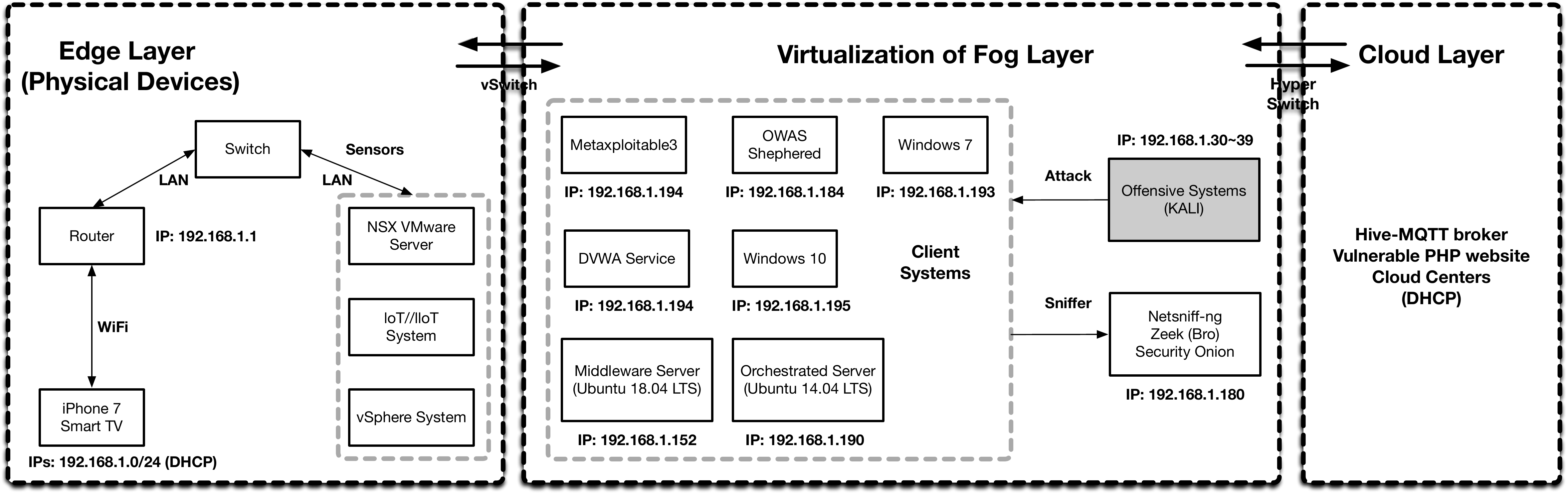}
    \caption{Ton\_IoT Cyber Threat Assessment Testbed}
    \label{fig:testbed}
\end{figure*}

\subsubsection{Data-Centric-Based Artificial Intelligence Defense System}

The weakness of signature-based IDS has become particularly apparent since the increasing complexity and number of unknown cyber attacks in recent years.
It is suggested that traditional rule-based detection engines should be replaced with artificial intelligence based engines to enhance the capability of IDS on discovering unknown attacks.

Unsupervised learning \cite{barlow1989unsupervised} and statistical learning, such as K-means\cite{krishna1999genetic} and Local Outlier Factor (LOF)\cite{kriegel2009loop}, are used for an initial attempt of developing AI for intrusion detection.
These approaches can construct a profile for prediction that is only based on normal activities but without requiring any previous knowledge of attack. 
The advantage of such approaches is able to significantly reduce the cost of data collection and data labelling; also, it can theoretically improve the possibility of discovering unknown attacks.
However, the argument is that unsupervised learning for intrusion detection may achieve an impressive performance in a controlled laboratory environment but it may demonstrate a significantly performance degradation in practice\cite{cs259d}.

Supervised learning is another AI approach used for intrusion detection, which has shown a great potential on practical implementations\cite{suaboot2020taxonomy}.
In contrast to unsupervised learning, supervised learning requires to learn a labelled dataset, which consists of both knowledge of normal and malicious activities.
The advantage of using labelled dataset is able to significantly narrow the scale of detection surface to avoid the \textit{\textbf{`Needles in Haystacks'}} problem\cite{sommer2010outside} that we have mentioned in Section~\ref{introduction}.
Main stream supervised learning algorithms can be classified into classical machine learning (ML) approaches, such as Support Vector Machine (SVM), Random Forest (RF) and Adaptive Boosting (AdaBoost), and deep learning (DL) approaches, such as Convolution Neural Network (CNN), Recurrent Neural Network (RNN) and Graph Neural Network (GNN).

From the algorithmic perspective, both classical and deep learning approaches have advantages and drawbacks.
Classical ML requires a series of well-defined features that are manually designed by human experts as inputs for training to achieve its effectiveness.   
The advantage of classical ML is that hand-designed features can improve the interpretability of ML for intrusion detection so that can further eliminate the semantic gap between AI detectors with security analysts. 
However, it is similar to the weakness of signature-based approach; these features are defined excessively specific, which may lead to a drop of generalization performance once the targeted attack category beginning to variate.

DL approaches have partially addressed the problem of feature dependency\cite{bengio2006curse} that confused the ML community in a decade.
DL utilizes neural networks to construct a stack of learning layers as fixed feature extractors, which allows new representations can be self-learnt from its old representations.
It has been proven that learnt features are better than hand-designed features for the task of generalized learning\cite{lecun2015deep}.
However, the drawback of DL approach is relatively contradictory to its advantage, because the features learnt by a neural network contain a certain degree of abstracted implications, which are hard to be understood, explained and analyzed by security analysts.
The problem makes AI experts have to select more explicable security dataset for learning and attempt to map the low-level meanings of abstracted representation to their corresponding high-level attack stages, such as which part in the feature map is on behalf of initial compromise, latency movement or exfiltration, which significantly limits the effectiveness and practical applicability on applying DL for intrusion detection.

\section{Ton\_IoT Cyber Threat Assessment Testbed}

Ton\_IoT cyber threat assessment testbed is built at the Cyber Range and IoT Lab, School of Engineering and Information technology (SEIT), UNSW Canberra located with the Australian Defence Force Academy (ADFA).
The testbed can simulate several cyber-attack events by using different hacking techniques to against web applications, IoT gateways and computer systems across the edge layer, fog layer and cloud layer as shown in Fig~\ref{fig:testbed}.  

Edge layer involves the physical devices and their operating systems utilized as the infrastructure of configuring the virtualization technology and cloud services at the layers of fog  and  cloud,  respectively. 
It includes multiple IoT devices, such as Modbus and light bulb sensors, smartphones and smart TVs, as well as host systems, such as workstations and servers, used to connect IoT devices, hypervisors and physical gateways (i.e. routers and switches) to the internet.
The hypervisor technology of NSX-VMware \cite{NSX-VMware} was installed on a host server at the edge layer to manage the Virtual Machines (VMs) created at the fog layer.

Fog layer includes the virtualization technology that controls the VMs and their services using the NSX-VMware and vCloud platform \cite{vCloudplatform} to offers the framework of executing SDN and NFV in the proposed testbed.
The NSX vCloud NFV platform enables the design of a dynamic testbed IoT network of the Ton\_IoT with creating and controlling several VMs for hacking and normal operations, allowing the communications between the edge, fog and cloud layers via vSwitches and  gateways.
This layer includes the nodes of virtual machines configured to generate the datasets, as explained below: 
\begin{itemize}
    \item \textbf{Orchestrated Server -} is one of the main virtualized servers configured in the testbed using the Ubuntu14.04 LTS with the IP address 192.168.1.190.
    This server offered many orchestrated services, such as FTP, Kerberos, HTTPs, and DNS to simulate real production networks and generate more simulated network traffic using the Ostinato Traffic Generator \cite{OstinatoTrafficGenerator} that transmits traffic to other VMs in the testbed.
    \item \textbf{Middleware Server -} is the IoT virtualized server deployed in the testbed using the Ubuntu 18.04 with the IP address 192.168.1.152. 
    This server included the scripts that run IoT services through public and local MQTT gateways utilized in the testbed and linked with the cloud layer to subscribe and publish the telemetry data of IoT sensors.
    \item \textbf{Client Systems -} include a Windows 7 VM (IP address:192.168.1.193), Windows 10 VM (IP address:192.168.1.195), DVWA web service (IP address:192.168.1.192), OWASP security Sphered VM (IP address:192.168.1.184), Metaspoitable3 (IP address:192.168.194). 
    The two windows were used as the remote web interface of the node-red (IP address:192.168.1.152) and their network traffic and audit traces were logged. 
    The DVWA (Damn Vulnerable Web App) \cite{dvma} was utilized to make security vulnerabilities through web applications hacked using the virtualized offensive systems. 
    The OWASP security Sphered VM \cite{owasp} is an open-source platform that has many security vulnerabilities against mobile and web applications exploited using the offensive systems.
    In addition, the Metasploitable3 VM \cite{meta3} was deployed in the testbed to increase vulnerable fog nodes and hack them using various attacking techniques by the offensive systems.  
    \item \textbf{Offensive Systems -} include the Kali Linux VMs and scripts of hacking scenarios that exploit vulnerable systems in the testbed network. 
    Ten static IP addresses (i.e. 192.168.1.30-39) were employed in the testbed to launch attacking scenarios and breach vulnerable systems  either IoT services (client and public MQTT brokers and node-red IP), operating systems (i.e. Windows 7 and 10, and Ubuntu 14.04 LTS and 18.04 LTS), and network systems (i.e. IP addresses and open protocols of the VMs).
    \item \textbf{Data Logger System -} is to log network traffic of the ToN\_IoT datasets, the Security Onion VM \cite{securityonionvm} (IP address:192.168.1.180) was used to log network data from all the active systems in the testbed using a virtual mirror switch that forwards the entire network traffic to this VM without dropping any traffic.
    As shown in Figure 1, the netsniff-ng tool \cite{netsniff-ng} was used to capture the entire network packets from the entire systems in pcap formats without packet drops.
    The Zeek Network Security Monitor tool (previously named Bro) \cite{bro} was used to generate data features from the pcap files.  
\end{itemize}

Cloud layer contains the cloud services configured online in the testbed.
The fog and edge services connected with the public HIVE MQTT dashboard \cite{hivemqtt}, public PHP vulnerable website \cite{php}, cloud virtualization, and cloud data analytics services (e.g., Microsoft Azure or AWS).
The public HIVE MQTT dashboard enabled us to publish and subscribe to the telemetry data of IoT services via the configuration of the node-red tool. 
The public PHP vulnerable website used to launch injection hacking events against websites. The other cloud services were configured either in Microsoft Azure or AWS to transmit sensory data to the cloud and visualize their patterns.

\renewcommand{\baselinestretch}{1.6}
\begin{table*}[t]
\caption{Ground Truth of UNSW-NB15}
\begin{center}
\begin{tabular}[\linewidth]{|c||c|c|}
\hline
\textbf{Attack Category} &\textbf{Attack Reference} &\textbf{Attack Description}	\\
	\hline \hline
\multirow{2}{*}{Reconnaissance}	&	CVE 2002-0563 	&	Oracle 9iAS Dynamic Monitoring Services Anonymous Access Variant 2	\\
\cline{2-3}
		&	CVE 2002-0563  &	Oracle 9iAS Dynamic Monitoring Services Anonymous Access Variant 8		\\
	\hline
\multirow{3}{*}{Exploits}	&	CVE 2009-0227	&	Microsoft Office PowerPoint Legacy File Format Stack Overflow (POP3 Base64) 		\\
\cline{2-3}
		&	CVE 2008-4261	&	Internet Explorer - HTML Rendering Buffer Overflow		\\
\cline{2-3}
	 &	CVE 2006-4076	&	BerliOS Docpile:we access.inc.php INIT\_PATH Parameter PHP File Include		\\
	\hline
\multirow{2}{*}{Fuzzers}	&	NULL	&	Fuzzer: OSPF Database Description Packet: Basic 		\\
\cline{2-3}
		&	NULL	&	Fuzzer: HTTP GET Request Invalid URI		\\
\hline
\multirow{2}{*}{Generic}	&	CVE 2011-2748	&	McAfee SiteManager ActiveX Control ExportSiteList Buffer Overflow 	\\
\cline{2-3}
	&	CVE 2006-2934	&	Linux Kernel SMB Filesystem receive Transaction2\_vulnerability 		\\
\hline
\multirow{3}{*}{DoS}	&	CVE 2013-6449 	&	OpenSSL ssl\_get\_algorithm2 TLS Denial of Service	\\
\cline{2-3}
	&	CVE 2011-0475	&	Google Chrome PDF Viewer Use-After-Free (HTTP) 		\\
\cline{2-3}
	&	CVE 2007-1030	&	DNS Label Compression Recursion (zlip-1/UDP)		\\
\hline
Analysis &	NULL &	Analysis: IP Protocol Scan	\\
\hline
\multirow{3}{*}{Backdoor}		&	CVE 2014-2269 	&	Vtiger CRM Unauthenticated Password Reset	\\
\cline{2-3}
		&	CVE 2012-5159 	&	phpmyadmin 3.5.2.2 Backdoor Access and Code Execution		\\
\cline{2-3}
		&	CVE 2013-3585	&	Samsung DVR Authentication Bypass		\\
\hline
\multirow{3}{*}{Shellcode}		&	milw0rm-1308	&	Linux PPC Read Execute - core (UDP)	\\
\cline{2-3}
		&	NULL	&	Mac OS X PPC Bind Shell - metasploit (UDP)	\\
\cline{2-3}
	& NULL	&	 Windows x86 Execute Command - metasploit (UDP) Variant 2		\\
\hline
\multirow{3}{*}{Worms}		&	MSB MS02-039	&	Microsoft SQL Server Slammer/Saphire Worm	\\
\cline{2-3}
	&	CVE 2006-2492	&	Trojan.MDropper Word Document (http) Variant 2	\\
\cline{2-3}
	&  CVE 2005-1921	&	 Lupper.A XML-RPC Propogation Request Variant 13		\\
\hline
\end{tabular}
\label{tab:UNSW-NB15GT}
\end{center}
\end{table*}
\renewcommand{\baselinestretch}{1}

\section{Description of Threat Model}

\subsection{Attack Scenarios on Ton\_IoT Testbed}
Ton\_IoT testbed contains several attack scenarios against vulnerable elements of IoT applications, operating systems, network systems, which are used to evaluate the proposed Densely-ResNet.
The scripts and some links of the attacking categories are published in \cite{TonIoTAttack}. 
The attack families utilized in the Ton\_IoT dataset are explained as follows: 
\begin{itemize}
    \item \textbf{Scanning attack -} we used the Nessus and Nmap tools from the offensive systems with IP addresses 192.168.1.20-38 against the target subnet 192.168.1.0/24 and all other public vulnerable systems such as the Public MQTT broker and vulnerable PHP website. For example, nmap 192.168.1.40-254, and the scans of the Nessus tool for the same range of IP addresses. 
    \item \textbf{Denial of Service (DoS) attack -} we utilized DoS attack scenarios on the offensive systems with IP addresses 192.168.1.\{30,31,39\} to hack vulnerable elements in the IoT testbed network. We created Python scripts using the Scapy package to launch the DoS attacks \cite{dosattack}.
    \item \textbf{Distributed Denial of Service (DDoS) attack -} we used DDoS attacks in the offensive systems wit IP addresses 192.168.1.\{30,31,34,35,36,37,38\} to breach several weaknesses in the IoT testbed network. We developed Python scripts using the Scapy package to launch the DoS attacks. 
    Further, automated bash scripts were developed to launch DDoS against vulnerable nodes of the testbed using the ufonet toolkit \cite{Ufonettoolkit}.
    \item \textbf{Ransomware attack -} we utilized the Kali Linux with IP addresses 192.168.1.\{33,37\} to execute this malware against windows operating systems and their webpages of monitoring IoT services included in the testbed network. 
    This attack executed using the Metasploit framework that hacks the SMB vulnerability of the systems, named eternalblue \cite{blue}. 
    \item \textbf{Backdoor attack -} we used the offensive systems with IP addresses 192.168.1.\{33,37\} to keep the hacking persistence using the Metasploit framework by executing a bash script of the command `run persistence -h' \cite{thebackdoorattack}.
    \item \textbf{Injection attack -} we used various injection scenarios from the offensive systems with IP addresses 192.168.1.\{30,31,33,35\} to inject data inputs against web applications of DVWA and Security Shepherd VMs and webpages of IoT services through other VMs, including SQL injection, client-side injection,broken authentication and data management, and unintended data leakage.
    \item \textbf{XSS attack -} we employed the offensive systems with IP addresses 192.168.1.\{32,35,36,39\} to illegally inject web applications of DVWA and Security Shepherd VMs and webpages of IoT services through other VMs. 
    In these systems, we created malicious bash scripts of python codes to hack the web applications of the testbed network using the Cross-Site Scripter toolkit (named XSSer) \cite{xsstoolkit}.
    \item \textbf{Password attack -} we used the offensive systems with IP addresses 192.168.1.\{30,31,32,35,38\}.
    In these systems, the hydra \cite{hydra} and cewl \cite{cewl} toolkits were configured using automated bash scripts to concurrently launch password hacking scenarios against vulnerable nodes in the testbed.
    \item \textbf{Man-In-The-Middle (MITM) attack -} we utilized the offensive systems with IP addresses 192.168.1.\{31,34\} to launch various MITM scenarios in the testbed network.
    In the systems, we employed the Ettercap tool \cite{ettercap} to execute ARP spoofing, ICMP redirection, port stealing and DHCP spoofing.

\end{itemize}

\begin{figure*}[t]
    \centering
    \includegraphics[width=.95\linewidth]{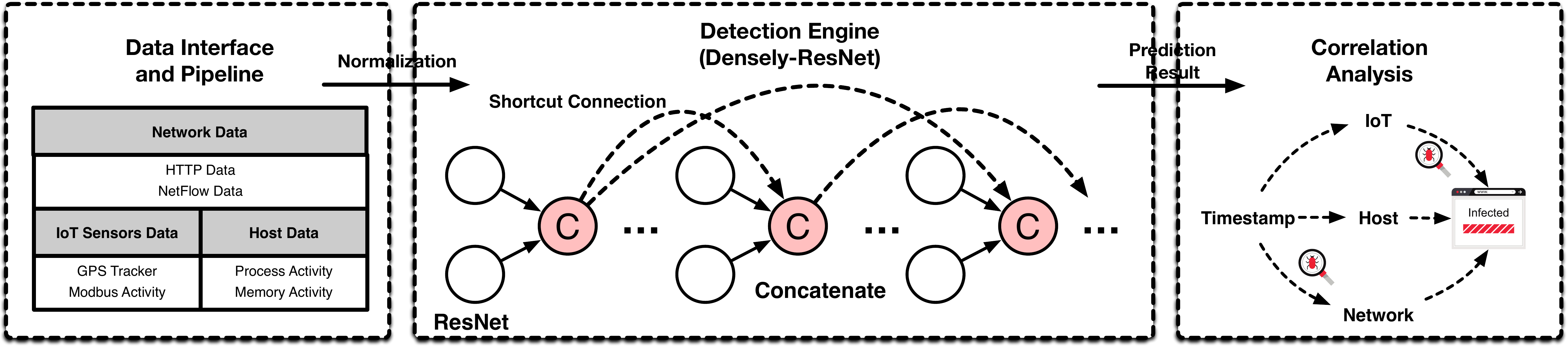}
    \caption{System Overview}
    \label{fig:system}
\end{figure*}

\subsection{Attack Scenarios on UNSW-NB15 Benchmark}
In addition to the Ton\_IoT testbed, we also evaluate the performance of Densely-ResNet on UNSW-NB15 benchmark dataset \cite{moustafa2015unsw,moustafa2016evaluation}.
The UNSW-NB15 dataset contains several attack categories: DoS, Exploits, Generic, Shellcode, Reconnaissance, Backdoor, Worms, Analysis and Fuzzers, which are collected from Common Vulnerabilities and Exposures\footnote{CVE: https://cve.mitre.org/}, Symantec\footnote{BID: https://www.securityfocus.com}, Microsoft Security Bulletin\footnote{MSD: https://docs.microsoft.com/en-us/security-updates/securitybulletins}.

The raw network packets of the UNSW-NB15 dataset is created by the IXIA PerfectStorm tool in the Cyber Range Lab of the Australian Centre for Cyber for generating a hybrid of real modern normal activities and synthetic contemporary attack behaviours.
It is worth noting that each attack event of UNSW-NB15 is simulated from a real-world attack scenario with a specific attack reference, which has been listed and described in Table 1.

Moreover, in our evaluation, we conduct the experiment with a plenty of real-world attack scenarios, which are included but are not only limited to those shown in Table 1. 
The actual attack reference used for our evaluation is in the range from CVE-1999-XXXX to CVE-2015-XXXX.

\section{Densely Residual Network (Densely-ResNet)}

Traditional ML suffers three main challenges for network intrusion detection:
\begin{enumerate}
    \item \textbf{Over-reliance on well-defined features}: Traditional ML requires specific and elaborate representations, which are usually designed by our security experts.
    However, hand-designed features include excessively detailed descriptions that commonly result in a poor generalization performance on discovering novel attacks.
    \item \textbf{Performance bottleneck}: Traditional ML, particularly kernel machine, suffers a problem called \textit{`the curse of dimensionality'}\cite{bengio2006curse}, which significantly limits the effectiveness of ML on learning big data with continuously increasing scale and complexity.
    \item \textbf{Unexplained result}: ML can contribute impressive result when training or testing data are under a well-curved distribution, such as a Gaussian Distribution. To achieve this, it is imperative to encode raw data to their statistical representations, which can be further used for a learning task.
    However, these statistical features are eventually inappropriate to be used for attack proof and identification for threat elimination.
\end{enumerate}

To address the above problems, we propose a densely connected residual network (Densely-ResNet) to reduce the cost of feature engineering, improve the generalization performance on discovering novel attacks and further enhance the result interpretability for attack proof and threat elimination.

\begin{figure}[htbp]
    \centering
    \includegraphics[width=.95\linewidth]{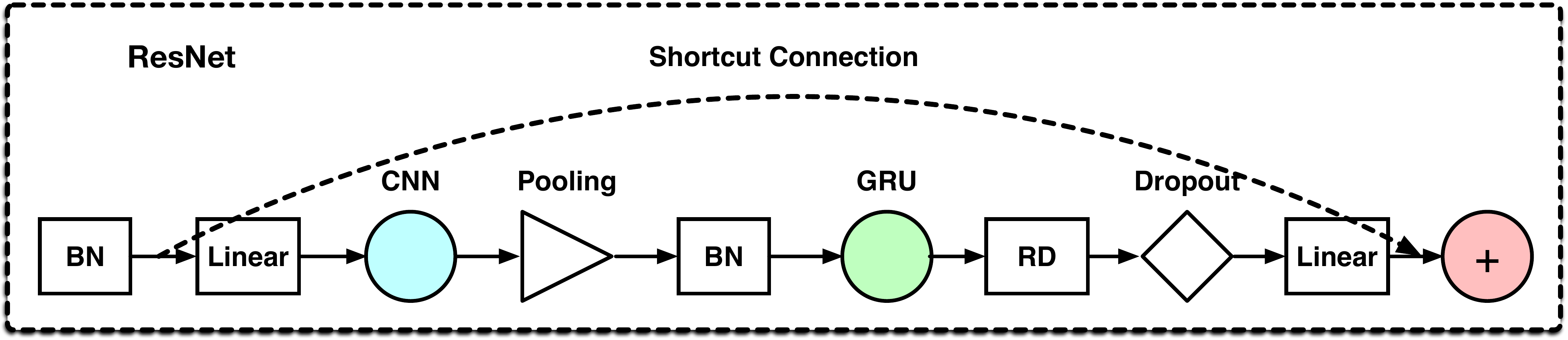}
    \caption{ResNet}
    \label{fig:ResNet}
\end{figure}

\subsection{System Overview}\label{AA}
Detection system consists of three modules: Data Interface and Pipeline, Detection Engine and Correlation Analysis; an overview of the whole detection system is shown in Fig~\ref{fig:system}.  

\textbf{Data Interface and Pipeline} is an entrance for multi-sourced data created through Edge, Fog and Cloud layers.
Since the multi-sourced data commonly have various formats, which cannot be directly processed by our detection engine, the module incorporates several data preprocessing functions to clean and normalize the multi-sourced data into a standard, readable and callable format. 

\textbf{Detection Engine} is the core of the detection system that is completely operated by Densely-ResNet.
Densely-ResNet is a customized deep neural network that consists of a series of ResNets, which are initially concatenated in a wide-channel direction and then eventually densely-connected from each anterior output to the rest of posterior outputs. 
Densely-ResNet overcomes several weaknesses of traditional neural networks and classical ML, such as gradient vanishing and dimension explosion.
Densely-ResNet is one of the most complex and effective known AI detectors for the task of network intrusion detection, which can achieve the state-of-the-art result on UNSW-NB15 benchmark in the evaluation.

\textbf{Correlation Analysis} is a post-analysis module for attack proof and identification. 
One problem is that events detected by our detection engine include a number of statistical contents, which are not explainable for an in-depth causality analysis.
Furthermore, another problem is that events alerted from a single layer or a single data source are not completely accurate.
The inaccurate alerts, however, are not always false positives, but should be considered as another situation, which is \textit{\textbf{`attack is happening but may not get success, in other words, the victim machine has not been compromised yet.'}}
The correlation analysis module is aimed to solve the above problems by integrating and analyzing alerts from Edge, Fog and Cloud layers in order to contribute more authentic and reliable detection results.

\subsection{Data Interface and Pipeline}
Data Interface and Pipeline is a socket that can convert and encode multi-sourced data to an unified format for the use of neural networks and other ML algorithms. 
The socket includes three functions as listed below:
\begin{itemize}
    \item \textbf{Content Normalization:} The function can convert multi-sourced data that have various types, such as xml, pcap and json, to an unified dataframe, which can be directely used for ML.
    Furthermore, the function can also discard some unexpected values, such as null values and duplicated records, which are harmful to the evaluation.
    \item \textbf{Feature Vectorization:} Multi-sourced data consist of many categorical features, such as ip address and protocol, which cannot be recognized by ML. The function uses a label-to-index strategy to encode these categorical features to vectors.
    \item \textbf{Standardization:} Data may have various distributions with different means and standard deviations, which significantly reduce the effectiveness and efficiency of ML. Hence, we use a standardization function to scale the data with a mean of 0 and a standard deviation of 1.
\end{itemize}

\subsection{Detection Engine (Densely-ResNet)}
Detection Engine uses a novel neural network architecture (Densely-ResNet) that we proposed in the paper.
Densely-ResNet is constructed by ten ResNets, where a pair of them is concatenated in a wide-channel direction as shown in Fig~\ref{fig:system}.
Each ResNet consists of a series of parameter layers as shown in Fig~\ref{fig:ResNet}, where BN is \textit{Batch Normalization}, CNN is \textit{Convolution Neural Network}, GRU is \textit{Gated Recurrent Unit} and RD is \textit{Reshape Dimensionality}.
Densely-ResNet has several remarkable advantages that enable to improve its generalization performance on discovering novel attacks.

\subsubsection{Spatial-temporal Learning}
It is a truth that CNN is effective to learn a grid of values, such as pixels in images, of which features are spatial-oriented; and GRU usually performs well on learning a series of sequence data, such as words and sentences, of which features are temporal-oriented.
However, for the multi-sourced security data, it is unclear to choose either CNN or GRU that can achieve better results.
Based on part of our experience and some prior work on network traffic analysis, we claim that, for most of one-dimension security data, it can be reasonably assumed that the data may have both spatial and temporal correlations.

\subsubsection{Feature Reuse}
Traditional neural network encounters a performance degradation problem when one would naively stack more layers onto the neural network for improving its learning capability.
The reason is that, when going deeper with neural network, learned features will gradually tend to be more specific, but far away from their original meanings, which eventually results in a problem of gradient vanishing.
To address the problem, Densely-ResNet adopts a strategy of feature reuse, which can mitigate the gradient loss both locally and globally.
To maximize the capability of feature reuse, as shown in Fig~\ref{fig:ResNet}, we uses a shortcut connection to add the output from one previous parameter layer to its subsequent parameter layer for keeping the local originality during the whole learning phase.
Moreover, as shown in Fig~\ref{fig:system}, each pair of the widely-connected ResNets will then be concatenated into all subsequent pairs of ResNets to keep the global originality during the whole learning phase.

\subsubsection{Linear Bridging}
Another problem of deep neural network is that, because of its non-linearity, the deep neural network will produce a slightly different validation result each time, which requires to be continuously optimized by retraining the network for obtaining the best result.
To reduce the cost of retraining a neural network, in Densely-ResNet, we use a Linear Bridging strategy that can transform a series of nonlinearized parameter layers into a linear space to improve the learning stability.
As a result, the final validation result will produce a extremely slighted fluctuation, which is not required to be retrained.

Densely-ResNet uses the global average pooling to finalise the whole learning phase and then amplifying the distinctions from selected active neurons.
In the end, Densely-ResNet uses a loss function of categorical cross-entropy to minimize the error rate for the task of multi-class classification.

\subsection{Correlation Analysis}

The correlation analysis module aims to make a causality analysis based on the provenance of attacks to identify those truly and successfully happened attack events from a large number of false positives and failed attempts.
To achieve this, the module will search for and take into considerations of the prediction results from our detection engine that happened concurrently in Edge, Fog and Cloud layers within a defined time threshold to address the problem of \textbf{\textit{`attacker knocked the door but was not there.'}}
We claim that a true intrusion event should be able to complete its attack actions in a short period and can be captured through correlating the detection results from user, host and network behaviour. 
Furthermore, the accuracy of the correlation analysis module should not be sensitive to the different selections of defined time thresholds, which will be discussed in our evaluation.

\renewcommand{\baselinestretch}{1.5}
\begin{table*}[t]
\caption{Testing Performance on Ton\_IoT Testbed}
\begin{center}
\begin{tabular}[\linewidth]{|c||c||c|c|c|c|c|c|c|c|c|c|}
\hline
\textbf{Layer} & \textbf{Type} & \textbf{ACC\%} & \textbf{TP} & \textbf{TN} & \textbf{FP}& \textbf{FN} & \textbf{DR\%}& \textbf{FAR\%}& \textbf{Precision\%}& \textbf{Recall\%}& \textbf{F1 Score}\\
\hline \hline
Edge & IoT & 98.27 & 15,611 & 24,500 & 0 & 0 & 100.00 & 0.00 & 100.00 & 100.00 & 100.00\\
\hline
\multirow{2}{*}{Fog} & Windows & 99.92 & 1,707 & 1,999 & 1 & 1 & 99.94 & 0.05 & 99.94 & 99.94 & 99.94\\
\cline{2-12}
& Linux & 95.51 & 15,870 & 29,940 & 60 & 163 & 98.98 & 0.20 & 99.62 & 98.98 & 99.30\\
\hline
Cloud & Network & 99.93 & 16,102 & 29,971 & 29 & 2 & 99.99 & 0.10 & 99.82 & 99.99 & 99.90\\
\hline
\end{tabular}
\label{tab:Testing Performance on Ton_IoT}
\end{center}
\end{table*}

\renewcommand{\baselinestretch}{1}

\section{Evaluation}

Our evaluation is based on a cloud AI platform configured with a Tesla K80 GPU and a total of 12 GB of RAM. 
The detection system is completely written in Python; and the detection engine is built upon tensorflow backend with the APIs of keras and scikit-learn packages.

\subsection{Feature Selection and Preprocessing}

There are 1,359,589 data records from Ton\_IoT testbed and 257,673 data records from UNSW-NB15 benchmark used for the evaluation.
To ensure the effectiveness of the evaluation, we select 10\% data records from the Ton\_IoT as a testing set, where the testing samples are never used for training.
For the UNSW-NB15 benchmark, we use the default training and testing sets for the evaluation, where the testing set is also totally unseen in the training set.

Selected features for Ton\_IoT are sensors activities from the Edge layer, CPU, memory, disk and process activities from the Fog layer, and network activities from the Cloud layer, whereas for UNSW-NB15 benchmark, there are only network activities available for use.
A detailed description of feature selection for the two datasets can be found in the referred articles\cite{TonIoTAttack,moustafa2015unsw}.
Finally, all data records used for both training and testing are uniformly processed by the Data Interface and Pipeline module. 

\subsection{Hyper-parameter Setting for Densely-ResNet }

Densely-ResNet requires to configure a series of hyper-parameters to initialize the deep neural network.
Filter size of convolution and number of recurrent unit must be equal and fixed to the number of features used for training in each dataset, where we set 42 for UNSW-NB15, 19 for IoT, 178 for Windows, 40 for Linux and 233 for network in Ton\_IoT respectively.
Furthermore, we claim that our model is insensitive to the other hyper-parameters, such as kernel size, dropout rate, which can be freely tuned in a reasonable manner.

\renewcommand{\baselinestretch}{1.6}
\begin{table*}[htbp]
\caption{Testing Performance on UNSW-NB15 Benchmark}
\begin{center}
\begin{tabular}[\linewidth]{|c||c|c|c|c|c|c|c|c|c|c|c|}
\hline
\textbf{Method} & \textbf{ACC\%} & \textbf{TP} & \textbf{TN} & \textbf{FP}& \textbf{FN} & \textbf{DR\%}& \textbf{FAR\%}& \textbf{Precision\%}& \textbf{Recall\%}& \textbf{F1 Score} &\textbf{Runtime/s}\\
\hline \hline
AdaBoost\cite{hu2013online} & 45.02 & 44,404 & 14,491 & 22,509 & 928 & 97.95 & 60.84 & 66.36 & 97.95 & 79.12 &40.42\\
\hline
RF\cite{zhang2008random} & 53.61 & 41,028 & 20,409 & 16,591 & 4,304 & 90.51 & 44.84 & 71.21 & 90.51 & 79.71 &7.49\\
\hline
SVM (RBF)\cite{ahmad2018performance} & 65.00 & 45,089 & 21,824 & 15,176 & 243 & 99.46 & 41.02 & 74.82 & 99.46 & 85.40 &2597.83\\
\hline
MLP\cite{pal1992multilayer} & 67.45 & 45,212 & 21,529 & 15,471 & 120 & 99.74 & 41.81 & 74.51 & 99.74 & 85.30 & 192.46\\
\hline
HAST-IDS\cite{wang2017hast} & 67.75 & 44,997 & 22,410 & 14,590 & 335 & 99.26 & 39.43 & 75.51 & 99.26 & 85.77 & 259.02\\
\hline
LSTM\cite{hochreiter1997long} & 68.27 & 44,083 & 24,231 & 12,769 & 1,249 & 97.24 & 34.51 & 77.54 & 97.24 & 86.28 & 133.35\\
\hline
CNN \cite{hinton2012neural}& 68.62 & 44,377 & 22,819 & 14,181 & 955 & 97.89 & 38.33 & 75.78 & 97.89 & 85.43 & 111.54\\
\hline
LuNet\cite{peilun} & 72.67 & 44,801 & 24,729 & 12,271 & 531 & 98.83 & 33.16 & 78.50 & 98.83 & 87.50 & 286.58\\
\hline
\textbf{Densely-ResNet} & \textbf{73.93} & \textbf{43,826} & \textbf{26,680} & \textbf{10,320} & \textbf{1,506} & \textbf{96.68} & \textbf{27.89} & \textbf{80.94} & \textbf{96.68} & \textbf{88.11} & \textbf{849.18}\\
\hline
\end{tabular}
\label{tab:Testing Performance on UNSW-NB15}
\end{center}
\end{table*}

\renewcommand{\baselinestretch}{1}

\subsection{Testing Performance on Ton\_IoT Testbed}\label{testTon}

Ton\_IoT include nine types of attacks for testing: Scanning, DoS, DDoS, Ransomware, Injection, Backdoor, XSS, Password and MITM.
An overall of testing performance of Densely-ResNet on Ton\_IoT testbed is shown in Table~\ref{tab:Testing Performance on Ton_IoT}.
Densely-ResNet is trained separately on the Edge, Fog and Cloud layers, where the Fog layer includes Windows and Linux endpoints, which also need to be trained separately.

For the Edge layer, the detection engine can 100\% detect all attacks monitored by IoT sensors, which include Backdoor, Password, Injection, DDoS, Ransomware, XSS and Scanning, without FARs.

For the Windows endpoint in Fog layer, the detection engine can achieve a 98.51\% DR with a 0.05\% FAR for Scanning attack, where the rest of attacks (DDoS, Password, Backdoor, Injection, XSS, DoS, Ransomware and MITM) can be 100\% detected without FARs.

For the Linux endpoint in Fog layer, the detection engine can achieve a 95.33\% DR for Password attack with a 0.19\% FAR, a 99.03\% DR for XSS attack and a 81.82\% DR for MITM attack without FARs, where the rest of attacks (DDoS, DoS, Injection, Scanning) can be 100\% detected without FARs.

For the Cloud layer, the detection engine can achieve a 99.95\% DR for DDoS attack without a FAR, a 99.95\% DR for XSS attack with a 0.03\% FAR, and a 100\% DR for Ransomware attack with a 0.06\% FAR, where the rest of attacks (Backdoor, Password, Injection, Scanning, DoS and MITM) can be 100\% detected without FARs.

\begin{figure}[htbp]
    \centering
    \includegraphics[width=.95\linewidth]{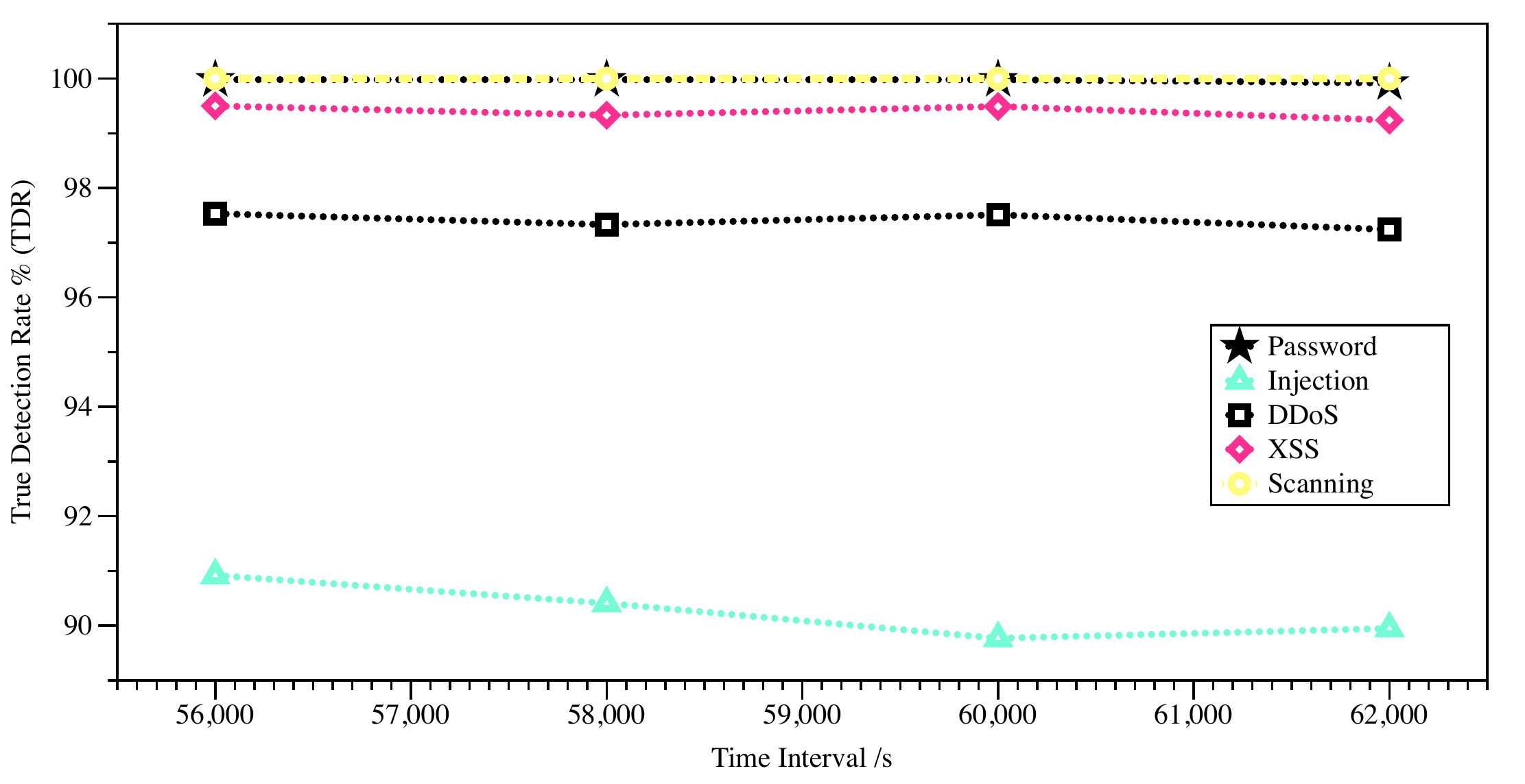}
    \caption{Correlation Analysis}
    \label{fig:correlation_analysis}
\end{figure}

\subsection{Correlation Analysis on Ton\_IoT Testbed}

The correlation analysis module is used to identify the attacks which are truly happened and caused consequences on the victim machine.
Due to most of attacks happened in different time periods, which cannot be completely captured and monitored, we select four different time intervals for the correlation analysis, which are 56000, 58000, 60000, 62000.
There are five attacks within the time intervals and concurrently happened in all Edge, Fog and Cloud layers, which are Password, Injection, DDoS, XSS and Scanning.
The result of correlation analysis is shown in Fig~\ref{fig:correlation_analysis}, where we use a metric of True Detection Rate (TDR) to measure the attacks simultaneously detected in the all three layers.
It can be observed that the TDR is lower than the DR that we presented in Section~\ref{testTon}, which means that a number of attacks happened but failed to compromise the victim machine.
Furthermore, it can also be observed that the TDR will not significantly fluctuate with the different interval of time windows.
Hence, we can confirm that the correlation analysis module is insensitive to the selection of time threshold.

\subsection{Result Interpretability}

\begin{figure}[htbp]
    \centering
    \includegraphics[width=.95\linewidth]{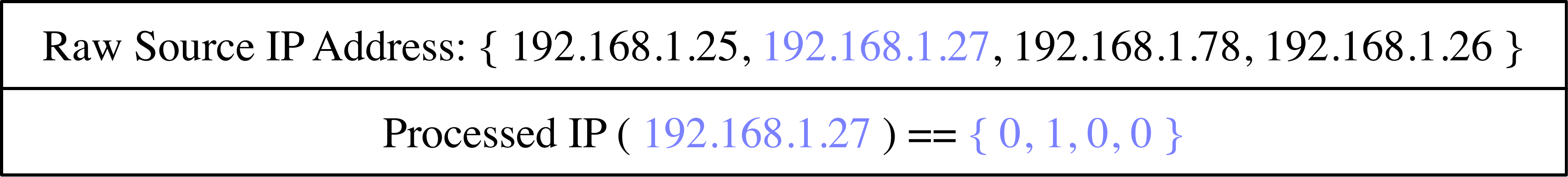}
    \caption{Result Interpretability}
    \label{fig:result_explain}
\end{figure}

Poor interpretability is one of the key problems on using ML for attack recognition.
The reason is that traditional ML cannot directly learn raw data, which include many categorical records with non-standard data formats, but requires to convert them as numeric values by using statistical methods, for example, ip address $192.168.1.27$ will be converted to $\{0,1,0,0\}$, which represents $True$ in four different existing ip records as shown in Fig~\ref{fig:result_explain}.

The converted features are eventually hard to be used for attack proof and post-analysis since they have lost the original meanings for explanation.
However, due to Densely-ResNet is built with several Conv-GRU structures, it is able to learn raw data records to keep their originality, which significantly improves the capability of result interpretability.

\subsection{Testing Performance on UNSW-NB15 Benchmark}

\begin{figure}[t]
    \centering
    \includegraphics[width=.95\linewidth]{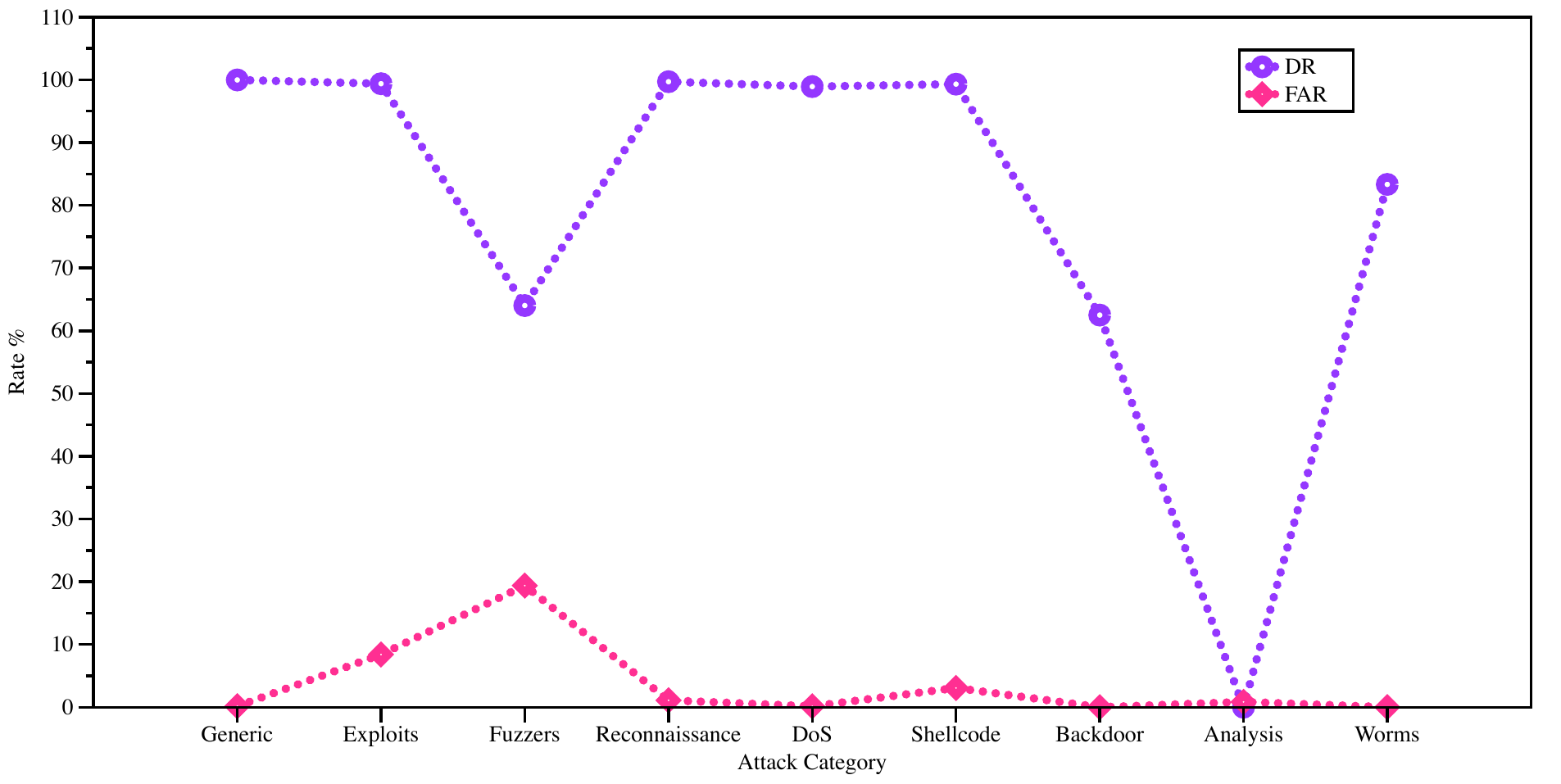}
    \caption{Detection Result on UNSW-NB15 Benchmark}
    \label{fig:unsw_atack_result}
\end{figure}

The testing set of UNSW-NB15 consists of around 86,000 unknown attacks that are never seen in the training set to make sure the reliability of the evaluation.
The detection result of Densely-ResNet on UNSW-NB15 benchmark is shown in Fig~\ref{fig:unsw_atack_result}.

With a closer inspection to the figure, Densely-ResNet can achieve an impressive detection performance on most of attacks. 
However, the detector has a moderate capability to detect Backdoor and Fuzzers and presents a higher FAR than others.
It is worth to note that our detection engine fails to detect the analysis attacks.
Tow reasons may lead to the poor detection result:
\begin{enumerate}
    \item Analysis behaviour are usually ambiguous with normal behaviour, such as executing some commands like $whoami$ and $ipconfig$, which are suspicious but may also be used by legitimated users.
    \item The records of analysis attack in UNSW-NB15 benchmark only take up around 0.27\%, which is an imbalanced learning problem that often results in a poor generalization performance.
\end{enumerate}

Furthermore, we also compare the generalization performance of Densely-ResNet with a number of ML and DL methods as shown in Table~\ref{tab:Testing Performance on UNSW-NB15}.
It is obvious that Densely-ResNet can achieve the best overall performance among these methods, particularly with the lowest FAR; however, we can also observe that Densely-ResNet leads to a higher run-time consumption, which is caused by its high algorithm and computation complexity.

\section{Conclusion}

In this paper, we introduce the Ton\_IoT testbed for cyber security assessment and propose the Densely-ResNet for intrusion detection that comes from Edge, Fog and Cloud layers.
We also conduct a correlation analysis to identify those attack events that have truly compromised the victim machine.

Furthermore, Densely-ResNet is evaluated on UNSW-NB15 benchmark to compare its generalization performance with a number of  ML and DL methods.
As a result, Densely-ResNet can achieve the state-of-the-art detection accuracy on UNSW-NB15 benchmark and simultaneously maintain a lower false alarm rate.
We confirm the effectiveness and robustness of Densely-ResNet based on the evaluation results and encourage researchers to use it for other tasks of intrusion detection in the future.

\section*{Acknowledgment}

Free use of the Ton\_IoT datasets for academic research purposes is hereby granted in perpetuity. Use for commercial purposes is allowable after asking the author, Dr Nour Moustafa, who has asserted his right under the Copyright. The datasets was sponsored by grants from the Australian Research Data Commons and UNSW Canberra.

\bibliographystyle{./bibliography/IEEEtran}
\bibliography{./bibliography/IEEEabrv,./bibliography/IEEEexample}

\end{document}